\documentclass[twocolumn, secnumarabic, amsmath, amssymb, nobibnotes, aps, prd, groupedaddress, superscriptaddress, nofootinbib]{revtex4-2}
\usepackage{multirow,slashed,hyperref,cleveref,ulem}
\usepackage{graphicx}
\usepackage{dcolumn}
\usepackage{bm}
\usepackage[dvipsnames]{xcolor}

\begin{document}
\title{Top partners and scalar dark matter -- a non-minimal reappraisal}

\preprint{KIAS-A22010}

\author{Alan S. Cornell}
\affiliation{Department of Physics, University of Johannesburg, PO Box 524, Auckland Park 2006, South Africa}
\email{acornell@uj.ac.za}
\author{Aldo Deandrea}
\affiliation{Universit\'e de Lyon, F-69622 Lyon, France: Universit\'e Lyon 1, Villeurbanne CNRS/IN2P3, UMR5822, Institut de Physique des 2 Infinis de Lyon}
\email{deandrea@ipnl.in2p3.fr}
\author{Thomas Flacke}
\affiliation{Center for AI and Natural Sciences, KIAS, Seoul 02455, Korea}
\email{flacke@kias.re.kr}
\author{Benjamin Fuks}
\affiliation{Laboratoire de Physique Th\'eorique et Hautes Energies (LPTHE), UMR 7589, Sorbonne Universit\'e et CNRS, 4 place Jussieu, 75252 Paris Cedex 05, France}
\email{fuks@lpthe.jussieu.fr}
\author{Lara Mason}
\affiliation{Department of Physics, University of Johannesburg, PO Box 524, Auckland Park 2006, South Africa}
\affiliation{Universit\'e de Lyon, F-69622 Lyon, France: Universit\'e Lyon 1, Villeurbanne CNRS/IN2P3, UMR5822, Institut de Physique des 2 Infinis de Lyon}
\email{mason@ipnl.in2p3.fr}

\begin{abstract}
A large set of models beyond the Standard Model of particle physics suggest that the top quark plays a special role in fundamental interactions. At the same time some of these models predict that a particle responsible for dark matter is present in the physical spectrum of the theory, and in particular cases dark matter may be specially linked to the top sector. Such a new physics configuration occurs, for instance, in composite frameworks that additionally feature vector-like quarks as mediators of the interactions of the dark matter with the Standard Model. We investigate the interplay of top-philic dark matter and the presence of several vector-like top partners. Such a non-minimal setup with several vector-like fermions is typical of composite scenarios, and we make use of an effective model to study to which extent such non-minimality modifies the collider and cosmological phenomenology of the minimal model, the latter only including one vector-like mediator of the dark matter interactions with the Standard Model.
\end{abstract}


\maketitle

\section{Introduction}
The idea that the top quark stands at a central point in the construction of models describing fundamental interactions is not a new one. In many models beyond the standard model (BSM) new sectors related to the top quark indeed play a prominent role. We explore the possibility that such interactions between the top quark and new physics arise from composite models (see~\cite{Panico:2015jxa,Cacciapaglia:2020kgq,Cacciapaglia:2022zwt} for recent reviews). Such frameworks could additionally predict the existence of a scalar (composite) heavy dark matter (DM) candidate $S$, that manifests as a weakly interacting massive particle coupling to the Standard Model (SM) top quark $t$ via a a Yukawa-type interaction involving a heavy vector-like fermionic mediator~$T$. 

Whilst we envisage the DM candidate $S$ and the vector-like fermionic mediator $T$ as emerging as bound states within a composite model with an underlying fermionic construction, BSM setups featuring fermionic mediators and bosonic top-philic dark matter are often used in effective models for dark matter~\cite{Zhang:2012da, Baek:2016lnv, Baek:2017ykw, Colucci:2018vxz, Biondini:2019int, Arina:2020udz, Cornell:2021crh, Belyaev:2022shr}. These include in particular vector-like portal models with scalar or vector dark matter. In these models, DM stability is guaranteed by a discrete $\mathbb{Z}_2$ symmetry under which the mediator and the dark boson are odd, whereas all SM fields are even. Corresponding composite UV completions however typically contain both new states of odd and even $\mathbb{Z}_2$ parity, and the particle spectrum includes in particular a new fermionic $T^\prime$ parity-even top partner. Whereas such a $\mathbb{Z}_2$-even $T'$ partner is usually present in composite Higgs models, it has been so far ignored when discussing the phenomenology relevant to top-philic simplified models for DM with a fermionic mediator. The purpose of the present work is to fill this gap, and to investigate to which extent following the path of next-to-minimality (by including the $T'$ state in the effective setup) impacts the phenomenology already captured by the minimal simplified model (with only $T$ and $S$ states included).

Our work thus further supports the possibility of heavy dark matter embedded in a composite Higgs scenario, but in contrast to existing literature, we consider the inclusion of an additional heavy fermionic resonance $T^\prime$ in the effective model. While minimal effective models only include the two $\mathbb{Z}_2$-odd states $S$ and $T$, our setup therefore involves not only the two $\mathbb{Z}_2$-odd states $S$ and $T$, but also the $\mathbb{Z}_2$-even state $T'$. We describe such a $T^\prime$ resonance as being a top partner, a standard inclusion within a generic composite Higgs model that facilitates the mechanism of partial compositeness~\cite{Kaplan:1991dc} by which fermions acquire mass. Although a composite Higgs model may feature a number of heavy vector-like quarks (VLQ), here we consider the inclusion of a single vector-like top partner $T'$ embedded in a generic representation of $SU(2)_L$, coupling to all Standard Model quarks. We recall that the name `vector-like' refers to the fact that the new fermions are not chiral with respect to the SM gauge interactions, and have therefore their left and right components lying within the same representation of the SM gauge group. 

In the rest of this letter, we look at the effects that the three states $S$, $T$ and $T'$ and their mutual interactions with the top quark have on collider phenomenology and cosmology. Our study is conducted relative to the phenomenological expectation of a reference minimal model that only features the $\mathbb{Z}_2$-odd states $S$ and $T$, and that has been already studied in the past~\cite{Colucci:2018vxz, Cornell:2021crh}. We follow the path introduced in~\cite{Buchkremer:2013bha, Fuks:2016ftf}, which proposes to start with a Lagrangian including all allowed couplings of the VLQs for model-independent search strategies. In section~\ref{sec:model}, we introduce this effective framework before focusing on the underlying cosmological implications and collider phenomenology in sections~\ref{sec:cosmo} and \ref{sec:colliders} respectively. We summarise our work in section~\ref{sec:concl}.

\section{An effective model}\label{sec:model}
In order to study the phenomenological implications of a next-to-minimal top-philic dark matter model that could emerge from a composite scenario, we consider a simplified model framework in which the SM is supplemented by two heavy fermionic top partners $T$ and $T'$ carrying the same quantum numbers as the right-handed top quark $t_R$, and a scalar dark matter candidate $S$. All new fields can be seen as bound states of some fundamental fermions, and the model is equipped with a $\mathbb{Z}_2$ symmetry to guarantee the stability of the DM particle. The $S$ and $T$ states are taken to be $\mathbb{Z}_2$-odd fields, whereas the $T'$ fermion is even. In realistic composite models with underlying fundamental (techni)-fermions the stability of the DM particle is often the result of a residual $U(1)$ symmetry. At the effective Lagrangian level, for symmetric $G/H$ coset spaces, a $\mathbb{Z}_2$ symmetry naturally emerges and can remain unbroken after introducing the matter fields. The charge under the $U(1)$ symmetry, or possible parities, depend on the details of the model. For a discussion on these points see for example \cite{Wu:2017iji,Csaki:2017jby}. The corresponding Lagrangian ${\cal L}$ reads
\begin{equation}
  \mathcal{L} = {\cal L}_{\rm SM} + {\cal L}_{\rm kin.} + \Big[
     \tilde{y}_t\          S\ \overline{T}\ t_R
   + \tilde{y}_{T^\prime}\ S\ \overline{T}\ T^\prime
   + {\rm H.c}.
   \Big]\,,
   \label{eq:tpintslag}
\end{equation}
where ${\cal L}_{\rm SM}$ is the SM Lagrangian, ${\cal L}_{\rm kin.}$ contains kinetic, mass and gauge interaction terms for all new fields (including the $T'$ interactions with the SM quarks and electroweak bosons that arise from the mixing induced by partial compositeness), and $\tilde{y}_t$ and $\tilde{y}_{T^\prime}$ represent the coupling strengths of the Yukawa interactions of the DM state with the top sector. 

The goal of this work is to assess the impact of the $\mathbb{Z}_2$-even state $T'$ on the model's phenomenology at colliders and in cosmology, this state being ignored in previous studies of similar models. Taking as a reference such a scenario in which the $T'$ is decoupled and irrelevant~\cite{Colucci:2018vxz}, we evaluate how existing constraints originating from the DM relic density, as well as its direct detection and indirect detection expectations, are modified by the presence of the $T'$ field. Particular emphasis is put on the interplay between the two Yukawa couplings $\tilde{y}_t$ and $\tilde{y}_{T^\prime}$ that may modify how DM generally scatters. Moreover, we also discuss $T’$-induced modifications to the collider phenomenology of the model, especially those that could emerge from the $\tilde{y}_t/\tilde{y}_{T^\prime}$ connection stemming from cosmology (a more visible and accessible collider signal leading to implications for DM direct detection, for instance). For scenarios exhibiting a decoupled $T'$ particle, existing (and updated) LHC constraints originating from Run~2 data have been determined, together with their projections for an integrated luminosity of 3 ab$^{-1}$, in \cite{Cornell:2021crh}. In this earlier work, results have been displayed in the two-mass plane $(m_S,m_T)$, the only Yukawa coupling of the model ($\tilde y_t$) being fixed in order to accommodate the observed DM relic density. It was shown that vector-like top partners with mass satisfying $m_T > 1.25$~TeV are excluded when the DM state $S$ is light, and that sensitivity is reduced for compressed mass spectra fulfilling $m_T\gtrsim m_S$. In addition, for scenarios with a decoupled $T'$ fermion and in which the $T'$ partner decays exclusively into $Wb$, $Zt$ and $ht$ systems, bounds from standard ATLAS and CMS searches for VLQs apply. Full Run~2 analyses~\cite{CMS:2022yxp, CMS:2022fck, ATLAS:2022ozf, ATLAS:2022tla, ATLAS:2022hnn} constrain viable scenarios to satisfy $m_{T'}\gtrsim 1.5$~TeV, the precise value depending on the hierarchy of the branching ratios of the $T'$ fermion. Constraints can nevertheless be substantially altered when non-standard decays of the $T'$ fermion are in order, {\it i.e.}\ when the $T’$ state can decay into non-SM particles. Such an option was comprehensively reviewed in \cite{Banerjee:2022xmu}, and is very relevant for the present work. Here, the $T'$ fermion is indeed coupled the $T-S$ system, the strength of this coupling being controlled by the size of the parameter $\tilde{y}_{T^\prime}$. Depending on the mass hierarchy between the $T$ and $T’$ partners and the DM particle $S$, additional handles on the model could be provided through collider signatures emerging from cascade decays of the top partners into each other. This is further addressed in section~\ref{sec:colliders}.

\section{Cosmological implications of the model}\label{sec:cosmo}
In this section, we estimate the impact of the $T'$ state and its interactions on cosmological observables, and in particular on those related to dark matter.
\begin{figure}
  \begin{center}
    \includegraphics[width=0.155\textwidth]{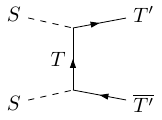}
	\includegraphics[width=0.155\textwidth]{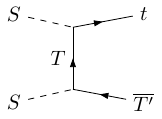}
	\includegraphics[width=0.155\textwidth]{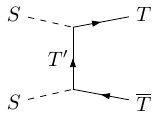}
  \end{center}
  \caption{Representative Feynman diagrams of contributions to DM annihilation that involve a top partner $T'$. Such diagrams either lead to the production of a $T'$ partner in the final state (first and second diagrams), or proceed via virtual $t$-channel $T'$ exchanges (third diagram).}\label{fig:tpams}
  \includegraphics[width=0.155\textwidth]{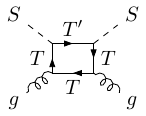}
  \includegraphics[width=0.155\textwidth]{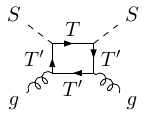}
  \includegraphics[width=0.155\textwidth]{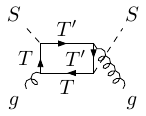}
  \caption{Representative one-loop Feynman diagrams of contributions to DM-nucleon scattering that involve a top partner $T'$, {\it i.e.}\ with at least one $T'$ state in the loop.}\label{fig:add_dd}
\end{figure}
The Lagrangian~\eqref{eq:tpintslag} allows for new contributions to the DM annihilation cross section in the early universe and today, relative to a reference scenario in which the $T'$ state is decoupled. A few examples are provided in Figure~\ref{fig:tpams}. $SS$ annihilation into a $T\overline{T}$ pair can be mediated via $t$-channel exchanges of $T'$ partners, whilst new annihilation channels into a $T'\overline{T'}$ pair or an associated $T'\bar{t}/t\overline{T'}$ pair can proceed through $t$-channel $T$ exchanges. The associated effects are studied in section~\ref{sec:relic}. In addition, we assess in section~\ref{sec:DMDD} the implications of the presence of the $T'$ partner in the particle spectrum on the DM direct detection cross section, as new loop diagram contributions of $t$, $T$ and $T'$ states to DM-nuclear scattering could be relevant. Examples of such contributions are shown in Figure~\ref{fig:add_dd}.

In our study, we consider a region of the parameter space in which the DM mass $m_S$ and the two top partner masses $m_T$ and $m_{T'}$ are within one order of magnitude of each other, keeping in mind their common composite origin. We then allow them to vary in the $[200, 3500]$~GeV range. Next, the two Yukawa couplings $\tilde{y}_t$ and $\tilde{y}_{T^\prime}$ are enforced to lie within the range $[10^{-4},6]$. This guarantees both a correct treatment of DM co-annihilations by {\sc MicrOMEGAS}~\cite{Belanger:2018ccd}, that we use for our numerical evaluations, and the validity of the adopted perturbative approach.

\subsection{DM relic density and indirect detection}\label{sec:relic}
In order to evaluate the DM relic density, we make use of the {\sc MicrOMEGAs} framework, in conjunction with a {\sc CalcHEP}~\cite{Belyaev:2012qa} model file generated with {\sc FeynRules}~\cite{Alloul:2013bka,Christensen:2009jx}. This allows for the automated treatment of all DM annihilation and co-annihilation channels, to which we add next-to-leading-order corrections in QCD that could be relevant in specific corners of the parameter space~\cite{Colucci:2018vxz,Colucci:2018qml}.

\begin{figure}
			\includegraphics[width=0.45\textwidth]{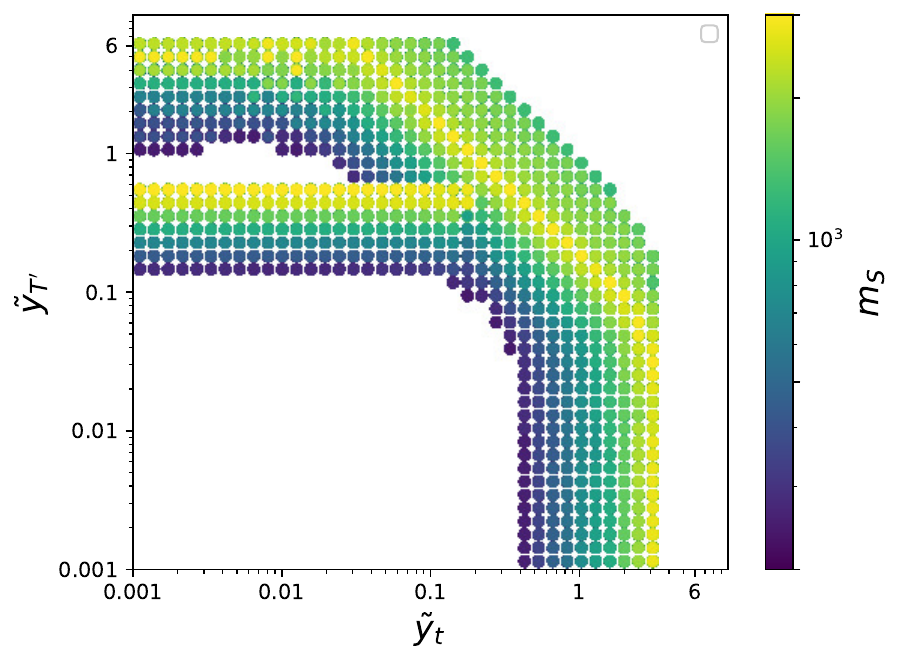}
		    \includegraphics[width=0.45\textwidth]{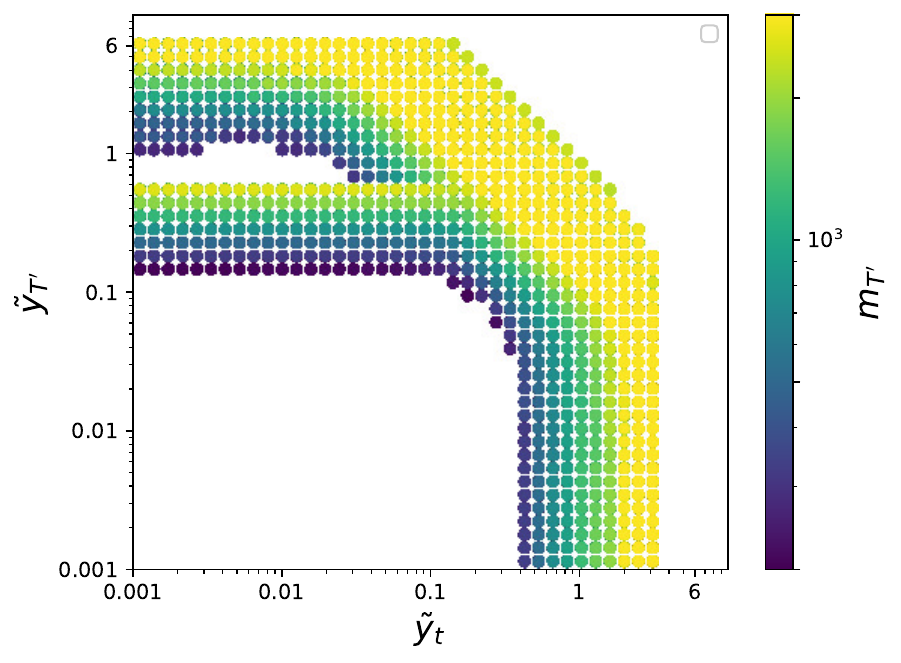}
		    \caption{Region of the parameter space in which a relic abundance of $\Omega_{\rm CDM}h^2 = 0.1186 \pm 0.0020$ can be accommodated, projected in the $(\tilde{y}_t, \tilde{y}_{T^\prime})$ plane. Each represented point corresponds to a Yukawa coupling configuration for which there exists at least one mass spectrum yielding the correct relic density. In the upper panel, we represent the DM mass $m_S$ corresponding to an allowed pair of Yukawa couplings through a colour code, whereas in the lower panel the colour code is related to the $T'$ mass $m_{T'}$. For $(\tilde{y}_t, \tilde{y}_{T^\prime})$ values for which several mass spectra exist, the colour code is associated with the largest possible mass.}
		    \label{fig:yukyuk}
\end{figure}

We perform a scan of our five-dimensional parameter space, and we retain the ensemble of scenarios for which predictions for the DM relic density $\Omega_{\rm CDM}h^2$ are compatible with the latest results from the Planck collaboration, $\Omega_{\rm CDM}h^2 = 0.1186 \pm 0.0020$~\cite{Aghanim:2018eyx}. The corresponding bounds are displayed in Figure~\ref{fig:yukyuk} in the $(\tilde{y}_t, \tilde{y}_{T^\prime})$ plane, the colour code indicating the mass of the dark matter (upper panel) or that of the $T'$ partner (lower panel). In these figures, all white regions correspond to coupling configurations for which there is no $S/T/T'$ mass spectrum yielding a relic density matching observations. 

Recalling that all new physics masses are within one order of magnitude of each other and taken between 200~GeV and 3500~GeV, the surviving scenarios feature either $m_{T'}<m_S<m_T$ or $m_S<m_{T'}, m_T$. These two possibilities manifest as two separate groups of allowed scenarios in the figures, especially visible at low $\tilde y_t$ values. For a given new physics mass hierarchy, the general scale of the new physics spectrum is correlated with the value of the $\tilde{y}_{T^\prime}$ coupling. Weaker $STT'$ interactions imply smaller new physics masses to get the right relic density, whereas stronger $STT'$ interactions require heavier states to compensate for the associated increase in the annihilation cross section.

We now compare our results with those obtained in the reference scenario in which the $T'$ state is decoupled~\cite{Colucci:2018vxz}. Such a decoupling can be obtained either by considering the lowest $\tilde{y}_{T^\prime}$ values scanned over, or through scenarios for which $2m_S < m_{T^\prime} + m_t$. In such a mass configuration, the $T'$ state is too heavy to affect the annihilation cross section obtained from the contributions that do not feature any $T'$ exchange or $T'$ partner in the final state. We observe that benchmarks with $\tilde{y}_t$ values smaller than about 1 can now be realised. In the reference scenario, $T$-mediated $SS\to t\bar t$ annihilations would not contribute strongly enough to the DM annihilation cross section for  $\tilde{y}_t < 0.3$. The presence of the $T'$ partner, however, enables a compensation of these too small contributions, and enhances the annihilation cross section via $T'$ exchange. This happens provided that the $STT'$ coupling strength is of ${\cal O}(1)$ or ${\cal O}(0.1)$ in the $m_S<m_{T'}, m_T$ and $m_{T'}<m_S<m_T$ hierarchy, respectively. In these cases, $\tilde{y}_t$ values as low as 0.0001 can be viable.

The effects depicted above affect the DM annihilation cross section at zero velocity in a similar way, and therefore the constraints originating from DM indirect detection. However, as in \cite{Colucci:2018vxz} for scenarios including a single $\mathbb{Z}_2$-odd top partner, resulting bounds are mild and only relevant for scenarios featuring a small DM mass of around 200 -- 300 GeV, and at least one large Yukawa coupling of ${\cal O}(1)$. As the focus on this work aims at describing the novelties that could emerge from a next-to-minimal scenario featuring two vector-like partners with different $\mathbb{Z}_2$ parity quantum numbers, we do not discuss indirect detection bounds further, and move on, in section~\ref{sec:DMDD}, with DM direct detection constraints for which large effects on the predictions have been found.

\subsection{DM direct detection constraints}\label{sec:DMDD}
As mentioned above, the Lagrangian~\eqref{eq:tpintslag} leads to contributions to DM scattering off nuclei through loops of top quarks, $T$ and $T'$ partners. These loop diagrams induce interactions between the DM state $S$ and the nucleon's gluonic content with a strength that depends both on the couplings appearing in~\eqref{eq:tpintslag}, and on the new physics mass spectrum. Using the corresponding expressions provided in~\cite{Hisano:2010ct,Hisano:2015bma} and numerical values  of~\cite{Hisano:2015bma} for the partonic mass fractions relating a gluon-DM coupling to a nucleon one, we deduce the spin-independent DM-nuclear and DM-proton scattering cross sections for each scenario for which the DM relic density constraints are satisfied.

\begin{figure}
			\includegraphics[width=0.45\textwidth]{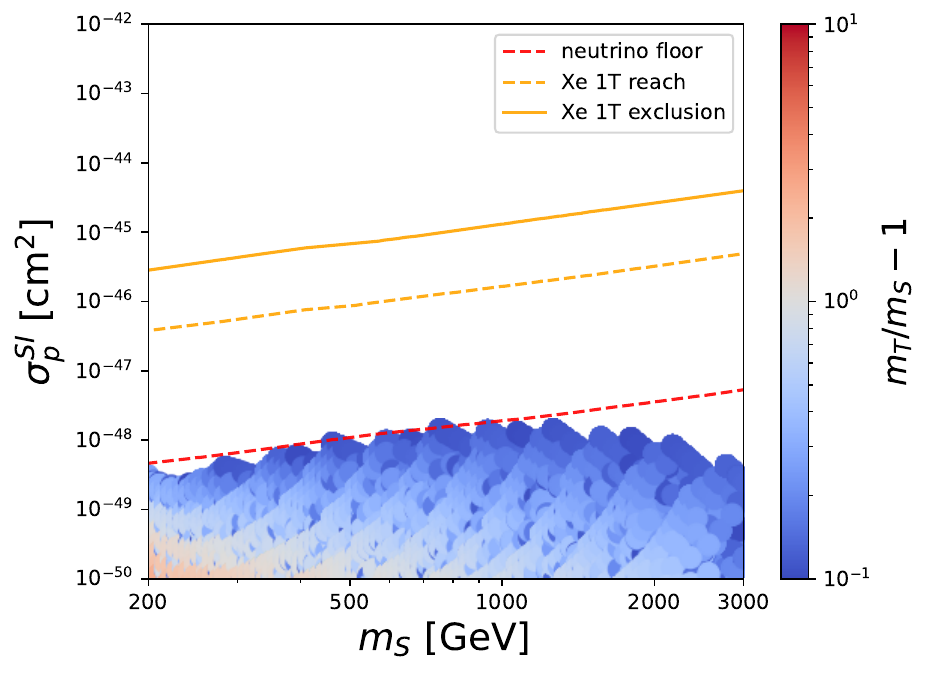}
		    \includegraphics[width=0.45\textwidth]{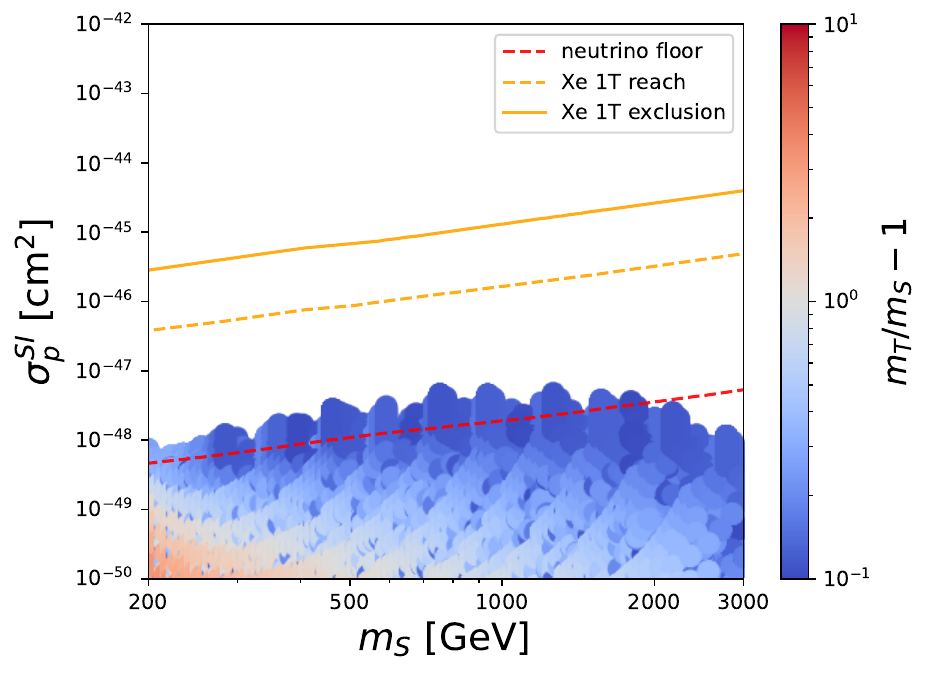}
		    \caption{DM-proton spin-independent scattering cross section as a function of the DM mass $m_S$, for two classes of scenarios. We consider new physics spectra in which the $\mathbb{Z}_2$-even $T'$ partner is decoupled (upper panel), or when its mass lies within one order of magnitude of the $S$ and $T$ masses (lower panel). All benchmarks are compatible with the DM relic density derived from Planck data. The colour code refers to the level of compression between the $T$ and $S$ states, and we additionally indicate the current and future sensitivity of the Xenon 1T experiment (orange lines), and the neutrino floor (red dashed line).}
		    \label{fig:ddTP}
\end{figure}

Our predictions are shown in Figure~\ref{fig:ddTP}, in which we estimate the impact of the $T'$ partner on the spin-independent DM-proton scattering cross section $\sigma_p^{SI}$. We display our results for two classes of scenarios. First, we decouple the $\mathbb{Z}_2$-even $T'$ partner by setting its mass to a large value and/or considering the smallest related Yukawa coupling values that we have scanned over. The corresponding predictions are shown in the upper panel of Figure~\ref{fig:ddTP}, together with information on the level of compression of the new particle mass spectrum through a colour code. 

As already remarked in previous studies~\cite{Colucci:2018vxz,Cornell:2021crh}, more compressed spectra lead to a smaller loop suppression, and therefore a larger scattering cross section. In addition, the cross section decreases with increasing values of the dark matter mass $m_S$, as stemming from the functional form of the DM-nuclear cross section. We superimpose in the figure the 90\% confidence level exclusion obtained by the Xenon 1T experiment after 34 days of exposure~\cite{Aprile:2017iyp} (solid orange line), and the Xenon 1T expected sensitivity under an assumption of 2.1 years of data acquisition~\cite{Aprile:2015uzo} (dashed orange line). Furthermore, we indicate through a dashed red line the neutrino floor~\cite{Billard:2013qya}, which allows for the classification of all tested scenarios in two categories, namely those potentially reachable at present and future DM direct detection experiments and those for which new experimental developments are needed. From our predictions, it is clear that a significant fraction of the scenarios compatible with the DM relic density constraints originating from Planck data will stay unconstrained in the near future by DM direct detection searches.

The situation changes significantly once the new particle spectrum includes a $T'$ state with a mass lying within at most one order of magnitude of the DM mass. The corresponding predictions are shown in the lower panel of Figure~\ref{fig:ddTP}. The additional loop-diagram contributions to the DM-nuclear scattering cross section lead to an enhancement of $\sigma_p^{SI}$ that ranges from a factor of a few to an order of magnitude. This is immediately visible through the comparison of the two sub-figures in Figure~\ref{fig:ddTP}. While the absolute cross section value stays very close to the neutrino floor (thus leading to poor DM direct detection experimental sensitivity), a non-negligible fraction of the scenarios returned by our scanning procedure yield predicted values for the DM-nuclear scattering cross section above the neutrino floor. Next-generation DM direct detection experiments have therefore a chance to be sensitive to these scenarios, and could thus be sensitive to the model and constrain it. 

In this subsection we have shown that making use of an effective theoretical framework with properties closer to those that could emerge from a UV-complete composite scenario leads to important modifications of the phenomenology of the model. The situation for the DM direct detection viewpoint on simplified models emerging from a composite framework is consequently less pessimistic than naively thought. Our direct detection results indeed demonstrate that it is important to leave a (minimal) simplified model description whenever possible (without increasing the complexity of the model too much), as the phenomenological implications may not be covered by the minimal case. Similar conclusions have been found, for instance, in supersymmetry~\cite{Fuks:2017rio} and in leptoquark scenarios attempting to explain the flavour anomalies~\cite{Borschensky:2022xsa}.

\section{Collider constraints on the model}\label{sec:colliders}

In the previous section, we investigated the cosmological impact of non-minimality on the effective top-philic $t$-channel dark matter model. Although non-minimality leads to complications in terms of the model's parameter space, it has the advantage of better describing the phenomenology of composite UV completions that could incorporate the initial effective set-up. In this section, we study how collider bounds are affected by the presence of both a $\mathbb{Z}_2$-odd ($T$) and a $\mathbb{Z}_2$-even ($T'$) top partner, as suggested in a composite model set-up. As in the previous section, we present our findings in comparison to a reference minimal scenario in which only a $\mathbb{Z}_2$-odd partner is active.

In such a minimal scenario and for proton-proton collisions at the LHC, new physics could manifest either through loop-induced mono-jet production, or from the production a pair of $T$ partners that both decay into DM and a top quark,
\begin{equation}\label{eq:processes1}
    p p \to S S j\quad\text{and}\quad p p \to T \bar T \to t S\ \bar t S \,.
\end{equation}
These signatures have been largely studied by the ATLAS and CMS collaborations, in particular in the context of simplified models inspired by the Minimal Supersymmetric Standard Model. Results relying on the full LHC run~2 dataset can be found, for instance, in~\cite{ATLAS:2019vcq, CMS:2020pyk}. Their re-interpretation in a minimal framework motivated by new strong dynamics in which the $T'$ state is decoupled demonstrated that scenarios compatible with cosmological bounds were further constrained by the results of the LHC~\cite{Cornell:2021crh}. For BSM configurations in which DM is light (with $m_S \sim 100~{\rm GeV}$), top partners with masses ranging up to about 1.25~TeV have been found to be excluded. Moreover, such a bound on $m_T$ only mildly changes for increasing DM mass values, the limits being of $1.15~{\rm TeV}$ for $m_S \sim 500~{\rm GeV}$, and of roughly $1.0~{\rm TeV}$ for $m_S \sim 700~{\rm GeV}$. Lighter $T$ options are nevertheless allowed for compressed setups, to which the LHC has generally less sensitivity ({\it i.e.}~for scenarios in which $m_T\approx m_S+m_t$).

We now move on with the analysis of the non-minimal setup considered. We investigate how the collider bounds derived in the minimal model are impacted by the presence of an additional $\mathbb{Z}_2$-even $T'$ partner with a mass close to that of the DM and $T$ states. In addition to the bounds previously discussed, which are specific to dark matter searches at the LHC, vector-like quark searches assuming the presence of the $T'$ state only must be accounted for. Consequently, strong constraints on the masses $m_S$, $m_{T'}$ and $m_T$ should be expected.

First of all, the $T'$ partner must have a mass that satisfies $m_{T'}\gtrsim 1.5~{\rm TeV}$ in order not to be in conflict with vector-like quark searches by the CMS and ATLAS collaborations (see~\cite{CMS:2022yxp,CMS:2022fck,ATLAS:2022ozf,ATLAS:2022tla,ATLAS:2022hnn} for analyses of the full run~2 LHC dataset), provided that the decay channel $T'\to TS$ is either negligible or at least sub-leading. Thus the bound on the $T'$ partner mass $m_{T'}$ could get slightly stronger than the one applicable on the mass of the $\mathbb{Z}_2$-odd $T$ partner $m_{T}$. Secondly, the presence of a $T'$ state could alter DM pair production in association with a jet through additional box diagrams involving both $T$ and $T'$ partners in the loop. However, it turns out that for the parameter space regions favoured by cosmology, these extra diagrams do not yield any substantial increase in the corresponding cross-section. Regardless of considering minimal or non-minimal models, mono-jet contributions to the DM signal at the LHC, arising from the $pp\to S S j$ process, remain negligible relative to those originating from the $pp\to T\bar T$ process (as in \cite{Colucci:2018vxz, Cornell:2021crh}).

It therefore remains to study how bounds stemming from searches for new physics in the production of a $t\bar{t}$ pair in association with missing transverse energy ($\slashed{E}_T$) are impacted by the presence of a $T'$ state, and how $T'$ bounds could be similarly weakened by the presence of the $T$ and $S$ particles in the spectrum. When $m_S<m_{T'}<m_T$ or $m_{T'}<m_S<m_T$, standard vector-like quark searches (for the $\mathbb{Z}_2$-even state $T'$) are unaffected by the new particles. The $\mathbb{Z}_2$-odd $T$ partner being heavier, the $T'$ state is not allowed to decay into a $TS$ system. The constraint {$m_{T'}\gtrsim 1.5~{\rm TeV}$ is therefore valid. On the contrary, when $m_S<m_T<m_{T'}$, bounds could in principle be modified. However, we argue in the following that such modifications are minor, and only relevant to a very small mass set of mass configurations $(m_T, m_{T'}, m_S)$.

If $m_S<m_T<m_{T'}<m_T+m_S$, there is no additional decay channel for the $T'$ partner, so that standard VLQ bounds are unaffected and apply. On the contrary, if $m_{T'}>m_T+m_S$ then the decay mode $T' \to T S$ opens up, and it could even be associated with a sizable branching fraction for large $\tilde{y}_{T^\prime}$ values (which concerns a wide class of scenarios motivated by cosmology, as shown in the previous section). However, the bounds on $m_T$ determined in previous works, namely $m_T > 1.15$~TeV for $m_S\lesssim 500$~GeV and $m_T > 1.25$~TeV for $m_S\lesssim 200$~GeV, remain valid. The VLQ bound $m_{T'} > 1.5$~TeV can thus only be relaxed in a small mass window in which a decay $T'\rightarrow T S$ is kinematically allowed while the existing bounds on $m_{T}$ are still satisfied. In those scenarios where a $T'\rightarrow T S$ decay is allowed, there are however new contributions to the $t \bar t + \slashed{E}_T$ signal, stemming from
\begin{equation}\label{eq:processes2}
    p p \rightarrow T'\bar{T'}\rightarrow T S\ \bar{T} S \rightarrow t S S\ \bar{t} S S\,.
\end{equation}
Consequently, the bound on $m_T$ gets more stringent, and the viable parameter space region in which $m_{T'}>m_T+m_S$ is further reduced. Studying precisely the modified bounds in this small mass window requires a recast of DM searches at the LHC in the $t\bar{t} + \slashed{E}_T$ mode~\cite{ATLAS:2019vcq, CMS:2020pyk}, as well as a recast of recent VLQ searches~\cite{CMS:2022yxp,CMS:2022fck,ATLAS:2022ozf,ATLAS:2022tla,ATLAS:2022hnn}. The former is available in public tools like {\sc MadAnalysis}~5~\cite{Conte:2018vmg}, and has in fact been used in \cite{Cornell:2021crh}. However, the latter is currently not available in any public tools. Performing associated implementations and validating them is highly non trivial~\cite{LHCReinterpretationForum:2020xtr}, and this lies beyond the scope of this letter. The reason is that the modifications of the bounds in the small relevant region of the parameter space (in which 1.5~TeV$>m_{T'}>m_T+m_S$ and $m_T>1.15$~TeV) are expected to be minor, and that the bounds are completely unaltered outside this mass regime. Investigating the constraints on our next-to-minimal new physics setup that were not present in the minimal case therefore reduces to the cosmological study of section~\ref{sec:cosmo}.

\section{Conclusion}\label{sec:concl}

Top-philic effective models for scalar dark matter typically rely on a simplified extension of the Standard Model with one dark scalar state, and one fermionic top partner playing the role of a mediator. To guarantee the stability of the dark matter, the two new states are taken to be odd under an {\it ad hoc} $\mathbb{Z}_2$ parity, all SM fields being of even parity quantum numbers. Such a framework can naturally be embedded in composite UV completions, in which both the dark matter and the mediator particles are bound states of fundamental fermions. It however turns out that other composite resonances are often close in mass to the dark matter and the mediator, and that they may be of $\mathbb{Z}_2$-even parity.

We have explored next-to-minimal effective scenarios for top-philic scalar dark matter, that include both $\mathbb{Z}_2$-odd and $\mathbb{Z}_2$-even top partners. Whereas the corresponding parameter space is more complicated than in the minimal setup, such a new physics configuration has the advantage of better describing any potential composite UV completion that features a vast zoo of new resonances. We have assessed the impact of this non-minimal case, relative to a minimal reference scenario in which all $\mathbb{Z}_2$-even new states are decoupled. 

It turns out that new parameter configurations can yield the correct relic density as extracted by the Planck collaboration, with possibly small new physics Yukawa couplings. Such a finding offers more viable parameter configurations than typically expected from the minimal model. In the latter case larger dark matter interaction strengths are required to accommodate the observed relic density. In addition, the presence of the $\mathbb{Z}_2$-even top partner may enhance the cross section associated with dark matter scattering off nuclei, making those scenarios potentially more likely to be found at future DM direct detection experiments. Such a conclusion improves the potential of such searches, and contrasts with the minimal setup in which the DM-proton scattering cross section is almost always below the neutrino floor. On the other hand, while the presence of the three new particles and their interactions lead to new contributions to collider probes of the model, their net impact is no different from a combination of the bounds originating from the minimal case with the additional constraints applicable on the $\mathbb{Z}_2$-even top partner. Light $\mathbb{Z}_2$-even top partners with masses smaller than about 1.5~TeV are excluded, and light $\mathbb{Z}_2$-odd top partners with masses below 1~TeV are excluded unless the spectrum features some level of compression, and this state becomes almost mass-degenerate with the dark matter candidate. One of the best options to probe deeper, from the currently viable region of the parameter space, is therefore that of present and future colliders, where the potential future impact of direct detection bounds are limited by the neutrino floor.

\section*{Acknowledgements}

This work was partly supported by funding from the {\it Cam\-pus-France} STAR project nr.~43566UG, as well as by the FKPPL LIA of the CNRS. ASC would like to thank the {\it Institut National de Physique} (INP) of the CNRS for funding his visits to LPTHE in Paris, and acknowledges partial support from the National Research Foundation in South Africa. The work of BF was supported in part by the French {\it Agence Nationale de la Recherche} (ANR) under grant ANR-21-CE31-0013 (project DMwithLLPatLHC). The work of TF is supported by a KIAS Individual Grant (AP083701) via the Center for AI and Natural Sciences at Korea Institute for Advanced Study.


\bibliographystyle{JHEP-2-2}
\bibliography{cdmbib}

\end{document}